\documentclass[10pt,twocolumn,twoside]{IEEEtran}

\usepackage{graphicx}
\usepackage{booktabs}
\usepackage{subfigure}
\usepackage{setspace}
\usepackage{multicol}
\usepackage{multirow}
\usepackage{epstopdf}
\usepackage{amsmath}
\usepackage{bm}
\usepackage{cite}
\usepackage{subfigure}
\usepackage{amssymb}
\usepackage{gensymb}
\usepackage{amsfonts}
\usepackage{mathrsfs}
\usepackage{amsmath}
\usepackage{algorithm}
\usepackage{algorithmic}
\usepackage{amsthm}

\usepackage{tabularx}
\usepackage{color}
\usepackage{balance}
\usepackage{mathrsfs}
\usepackage{setspace}
\usepackage{amsthm}
\usepackage{array}
\usepackage{cases} 
\usepackage{epstopdf} 

\newcommand{\PreserveBackslash}[1]{\let\temp=\\#1\let\\=\temp}
\newcolumntype{C}[1]{>{\PreserveBackslash\centering}p{#1}}
\usepackage[flushleft]{threeparttable}
\usepackage{setspace}

\usepackage{acronym}

\acrodef{snr}[SNR]{signal-to-noise ratio}%
\acrodef{ris}[RIS]{reconfigurable intelligent surface}
\acrodef{pin}[PIN]{positive-intrinsic-negative}
\acrodef{bs}[BS]{base station}
\acrodef{csi}[CSI]{channel state information}

\begin{document}


\title{Wavenumber-Domain Signal Processing for Holographic MIMO: Foundations, Methods, and Future Directions}
\author{ 
	Zijian Zhang,~\IEEEmembership{Graduate Student Member,~IEEE}, and Linglong Dai,~\IEEEmembership{Fellow,~IEEE}
	
	\thanks{Zijian Zhang and Linglong Dai are with the Department of Electronic Engineering, Tsinghua University, Beijing 100084, China, and also with the Beijing National Research Center for Information Science and Technology (BNRist), Beijing 100084, China. (e-mails: zhangzij15@tsinghua.org.cn; daill@tsinghua.edu.cn). The corresponding author is Linglong Dai.}
	
}
\maketitle
\IEEEpeerreviewmaketitle
\begin{abstract}	
Holographic multiple-input multiple-output (H-MIMO) systems represent a paradigm shift in wireless communications by enabling quasi-continuous apertures. Unlike conventional MIMO systems, H-MIMO with subwavelength antenna spacing operates in both far-field and near-field regimes, where classical discrete Fourier transform (DFT) representations fail to sufficiently capture the channel characteristics. To address this challenge, this article provides an overview of the emerging wavenumber-domain signal processing framework. Specifically, by leveraging spatial Fourier plane-wave decomposition to model H-MIMO channels, the wavenumber domain offers a unified and physically consistent basis for characterizing subwavelength-level spatial correlation and spherical wave propagation. This article first introduces the concept of H-MIMO and the wavenumber representation of H-MIMO channels. Next, it elaborates on wavenumber-domain signal processing technologies reported in the literature, including multiplexing, channel estimation, and waveform designs. Finally, it highlights open challenges and outlines future research directions in wavenumber-domain signal processing for next-generation wireless systems.
\end{abstract}
\section{Introduction}
The forthcoming sixth-generation (6G) wireless era requires unprecedented spectral and energy efficiency, ultra-low latency, and massive connectivity. To meet these demands, holographic multiple-input multiple-output (H-MIMO) communications have emerged as a promising technology \cite{HuangChongwen'20,gong2023holographic}. Specifically, H-MIMO envisions spatially quasi-continuous electromagnetic (EM) apertures enabled by metasurfaces. By densely integrating an extremely large number of sub-wavelength reconfigurable elements into a finite area, the quasi-continuous aperture of H-MIMO allows precise control of the amplitude, phase, and polarization of radiated EM waves \cite{AnJiancheng'CL'2023'I}. Such holographic apertures can approach the ultimate capacity limits of wireless channels \cite{ouyang2025diversity}. Moreover, as the array aperture grows electromagnetically large, the Rayleigh distance increases dramatically \cite{Cui'22'TCOM}, which motivates new representations for H-MIMO channels. By exploiting new channel properties, H-MIMO can open new spatial degrees of freedom (DoFs) to significantly enhance wireless communications. 

Conventional multiple-input multiple-output (MIMO) with half-wavelength antenna spacing usually relies on far-field channel models. In these settings, signals are often decomposed into a finite set of angular basis vectors via the discrete Fourier transform (DFT), which effectively captures propagation in the far-field region \cite{Cui'22'TCOM}. However, for H-MIMO with antenna-dense and large apertures, this assumption no longer holds. Firstly, the antenna spacing of H-MIMO is much smaller than half wavelength, which implies that the steering vectors of H-MIMO fundamentally differ from the DFT basis. Secondly, near-field array responses depend on both angle and distance, leading  to angle–distance coupling \cite{Cui'22'TCOM}. Due to the subwavelength-level spatial correlation and spherical wave propagation, classical DFT representations lose their sparsity and accuracy. In this context, the standard DFT representation suffers from “energy leakage”, causing significant performance degradation \cite{Cui'22'TCOM}. This phenomenon indicates that conventional signal processing methods are inadequate for H-MIMO, prompting the search for alternative frameworks.

Fortunately, the wavenumber domain offers a natural solution to this problem. Specifically, by expanding the EM field on holographic surfaces into a continuum of continuous plane-wave modes, i.e., Fourier harmonics, the wavenumber-domain representation inherently accounts for the subwavelength-level spatial correlation and the spherical propagation of EM waves. Thanks to its antenna-spacing-independent and distance-independent properties, the wavenumber domain is well-suitable for modeling and processing H-MIMO channels \cite{Yuanbin'25}. Recently, researchers have begun exploring wavenumber-domain signal processing. For example, the wavenumber-domain Fourier bases can serve as a waveform codebook for channel estimation \cite{guo2024wavenumber} or pattern designs \cite{Zijian'23'JSAC} in H-MIMO systems. The analysis from the wavenumber-domain perspective is also used in electromagnetic information theory (EIT) \cite{Jieao'24}, providing a foundational basis to find insightful results in H-MIMO communications.


Unlike existing works which primarily offer a system-level perspective \cite{HuangChongwen'20,gong2023holographic,AnJiancheng'CL'2023'I}, this article aims to provide a unified and physics-consistent treatment of H-MIMO from the wavenumber-domain viewpoint, emphasizing the role of plane-wave decomposition in signal processing. Specifically, in Section II, we introduce the foundations including the concept of H-MIMO and the wavenumber-domain representation of H-MIMO channels. In Section III, the existing wavenumber-domain signal processing are discussed, such as multiplexing, channel estimation, and waveform designs, with emphasis on their advantages and limitations. Then, Section IV discusses open challenges and promising research directions for the wavenumber-domain signal processing, such as the unified capacity theory, wideband communications, and network-level integration. Finally, conclusions are drawn in Section V.

\section{Foundations of Wavenumber-Domain Signal Processing for H-MIMO Systems}
In this section, we introduce the foundations of the wavenumber-domain signal processing for H-MIMO by clarifying the concept of H-MIMO and the wavenumber-domain representation of H-MIMO channels. We also discuss the opportunities in wavenumber-domain signal processing.

\subsection{Concept of H-MIMO}
\begin{figure}[!t]
	\centering
	\includegraphics[width=0.45\textwidth]{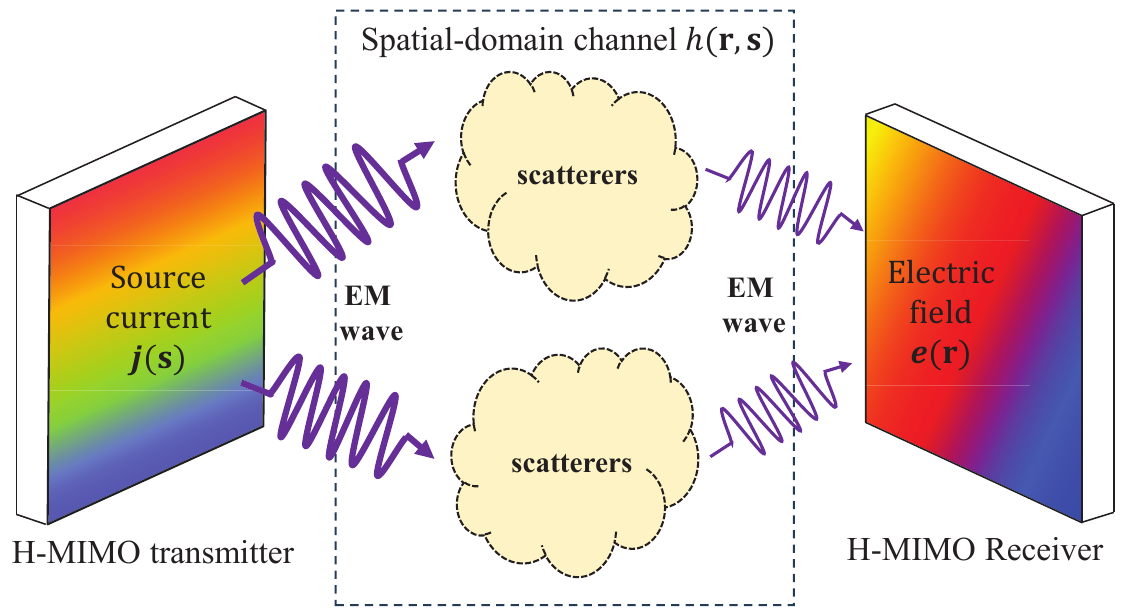}
	\caption{An illustration of H-MIMO based transmissions.}
	\label{img:H-MIMO}
\end{figure}
As shown in Fig. \ref{img:H-MIMO}, H-MIMO extends classical MIMO by leveraging metasurfaces or dense antenna arrays to approach a continuous EM aperture \cite{AnJiancheng'CL'2023'I}. By densely packing nearly infinite radiating elements into a finite area, one forms a spatially quasi-continuous EM aperture, which can approach the ultimate capacity limit of wireless channels \cite{Zijian'23'JSAC}. Such apertures can implement dynamic amplitude and phase control in the radio frequency (RF)-free domain at low cost. Crucially, an electromagnetically large aperture leads to an enlarged Rayleigh distance $2D^2/\lambda$, where $D$ is the array diameter and $\lambda$ is the wavelength. For sufficiently large $D$, e.g., tens of wavelengths, typical H-MIMO links will lie in the radiating near-field region rather than the far-field region. In the near-field, the array response exhibits angle-distance coupling, where the wavefront is spherical and EM-wave focus can be jointly controlled in angle and distance \cite{Cui'22'TCOM}. 

For conventional MIMO with $\lambda/2$ antenna spacing, one often assumes the far-field approximation. In that case, DFT bases suffice to discretize the spatial field. However, in H-MIMO systems, the subwavelength antenna spacing and the spherical nature of waves should be taken into account. The work \cite{Yuanbin'25} emphasizes that with an asymptotically continuous aperture, DFT-based channel representations cease to hold, and propagation should be modeled by exact EM wave equations. Thus, H-MIMO channels are more naturally expressed through their EM characteristics than by finite-path geometric models. To this end, H-MIMO channel models should consider (i) the spatially continuous aperture and (ii) the electromagnetically large aperture size. The former implies that channels can be modeled by surface current distributions, and the latter means that spatial correlation and mutual coupling (MC) cannot be neglected. As a result, H-MIMO channels are fundamentally EM channels, which requires physics-based modeling.

\subsection{Wavenumber-Domain H-MIMO Channels}
Generally, the wavenumber domain is a way of representing spatial waves as a sum of plane-wave components indexed by their wavenumbers, which describe how rapidly the wave oscillates across space. To analyze the physical channels of H-MIMO, the wavenumber-domain representation expresses the EM field on an H-MIMO aperture as a superposition of Fourier plane waves characterized by wavevectors ${\bf k}=[k_x,k_y,k_z]$, with $\|{\bf k}\|=2\pi/\lambda$. The key insight is that spherical waves can be expanded into a continuum of plane waves \cite{Pizzo'22}. Due to the fact that this representation defines continuous wavenumber-domain channels from the physical perspective, it is antenna-spacing-independent and distance-invariant, which can naturally describe the subwavelength-level spatial correlation and spherical-wave propagation in H-MIMO systems.

\begin{figure}[!t]
	\centering
	\includegraphics[width=0.45\textwidth]{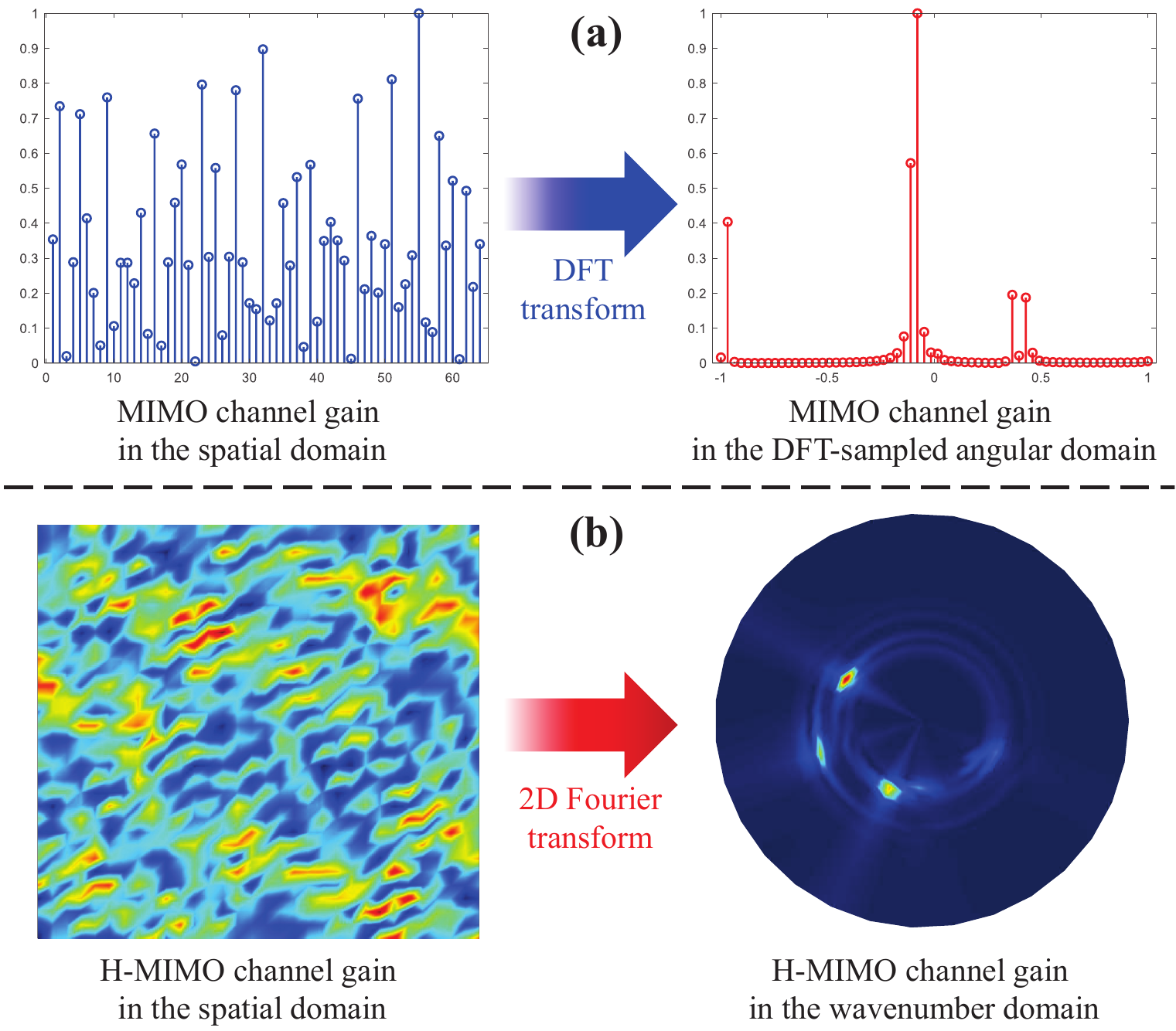}
	\caption{(a) The channel gains of a 64-antenna MIMO in the spatial domain and the DFT-sampled angular domain. (b) The channel gains of a continuous-aperture H-MIMO in the spatial domain and the wavenumber domain.}
	\label{img:channel}
\end{figure}

It is worth noting that, although both the DFT codebook and the wavenumber-domain representation rely on Fourier principles, they differ fundamentally in terms of physical assumptions, domain definitions, and applicability. The classical DFT basis assumes far-field propagation, $\lambda/2$ antenna spacing, and a discrete angular grid tied to the number of antennas. In contrast, the wavenumber-domain basis is derived from the spatial Fourier decomposition of EM fields, providing a continuous transverse-wavenumber spectrum, which naturally describes spherical-wave propagation, angle–distance coupling, and evanescent components. Consequently, the wavenumber domain remains valid for near-field and scenarios with sub-wavelength antenna spacing, where DFT-based representations suffer from strong basis mismatch and energy leakage. Therefore, the wavenumber-domain approach should not be regarded as a subset of the DFT codebook, but rather as a physically grounded basis essential for accurate modeling and processing in both H-MIMO and classical MIMO systems.

Particularly, existing work \cite{Pizzo'22} showed that, one can define a Fourier basis that captures the H-MIMO channel response across both near field and far field. In this way, each basis function corresponds to a plane wave of a given transverse wavenumber $(k_x,k_y)$. As illustrated in Fig. \ref{img:channel}, unlike a finite DFT basis which assumes plane waves anchored at discretized angles, the wavenumber basis allows “continuous” angle representation. In \cite{Yuanbin'25}, the authors refer to this as the {\it wavenumber-domain basis} and show it exposes clustering sparsity in H-MIMO channels. In contrast to the DFT basis, the wavenumber basis is not tied to a fixed grid of angles and so does not suffer from basis mismatch in H-MIMO systems.


\subsection{Opportunities in Signal Processing}

\begin{figure}[!t]
	\centering
	\includegraphics[width=0.42\textwidth]{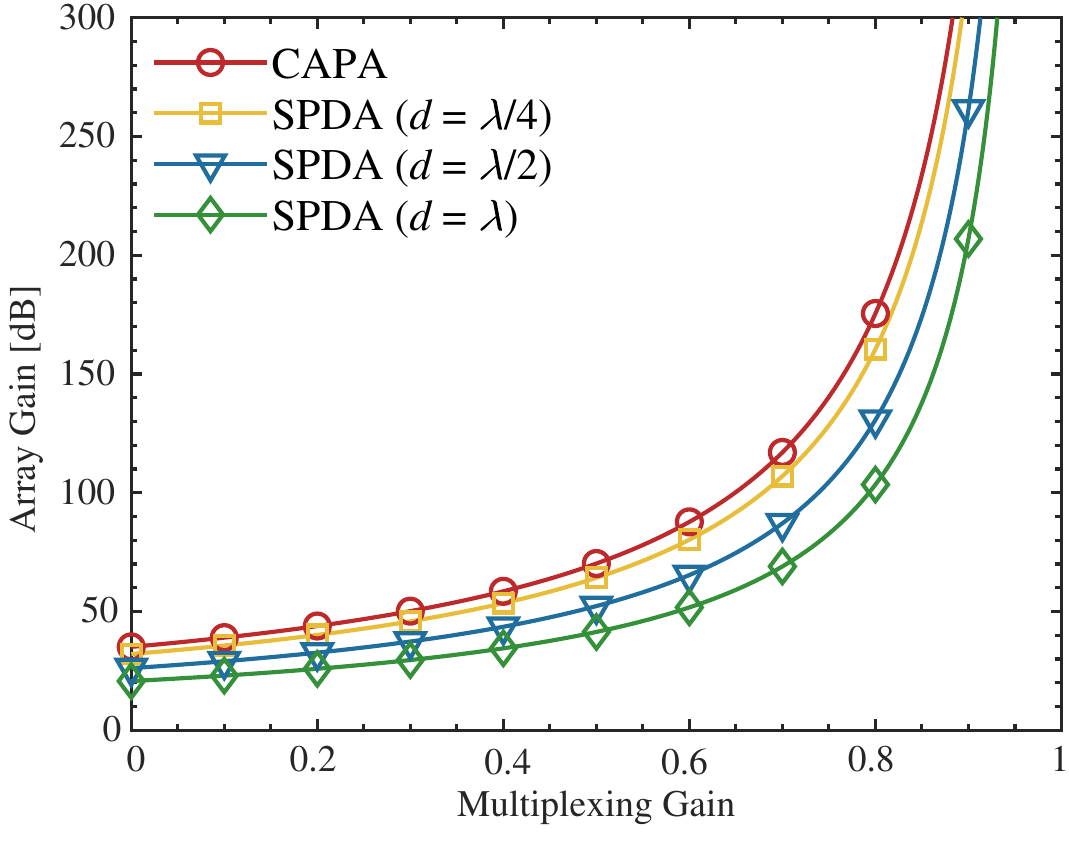}
	\caption{ The trade-off of array gain and multiplexing gain for CAPA and SPDA with different antenna spacing $d$ \cite{ouyang2025diversity}.}
	\label{img:DMT}
\end{figure}

The wavenumber-domain framework, grounded in EM theory, provides a unified approach for processing H-MIMO channels. It treats the transceiver apertures as continuous spatial domains and expands signals into plane-wave components, which provides opportunities for advanced signal processing. Particularly, the H-MIMO channel can be interpreted as a continuous linear mapping from the transmit aperture to the receive aperture, governed by Maxwell’s equations \cite{Jieao'24}. Building upon this framework, the trade-off between array gain and spatial multiplexing for continuous aperture arrays (CAPAs) and spatially discrete phased arrays (SPDAs) was analyzed in \cite{ouyang2025diversity}. As illustrated in Fig. \ref{img:DMT}, through proper waveform designs, H-MIMO systems employing CAPAs can achieve superior array gain or multiplexing performance compared to SPDAs. This analysis provides theoretical justification for the performance gains enabled by signal processing in H-MIMO systems.

Particularly, the wavenumber perspective naturally leads to spatial waveform designs. A transmitter can shape its aperture excitation so as to project power into specific wavenumber modes. For example, focusing a beam at a near-field point requires combining multiple plane-wave modes. Digital or hybrid precoding architectures can implement such wavenumber-selective beams, provided that the basis is known. The standard far-field strategy would be to use a DFT codebook, but this mismatches in H-MIMO systems. Instead, the authors in \cite{Yuanbin'25} proposed using the wavenumber-domain basis vectors as waveforming codewords. They showed that the wavenumber-domain designed waveformers can provide better performance and robustness than DFT codebook. In the next section, we will discuss the works on wavenumber-domain signal processing in the literature.

\section{Wavenumber-Domain Signal Processing}
Wavenumber-domain signal processing encompasses a wide range of techniques tailored for H-MIMO systems. In this section, we categorize existing designs into the following major groups.

\subsection{Wavenumber-Division Multiplexing}
\begin{figure}[!t]
	\centering
	\includegraphics[width=0.4\textwidth]{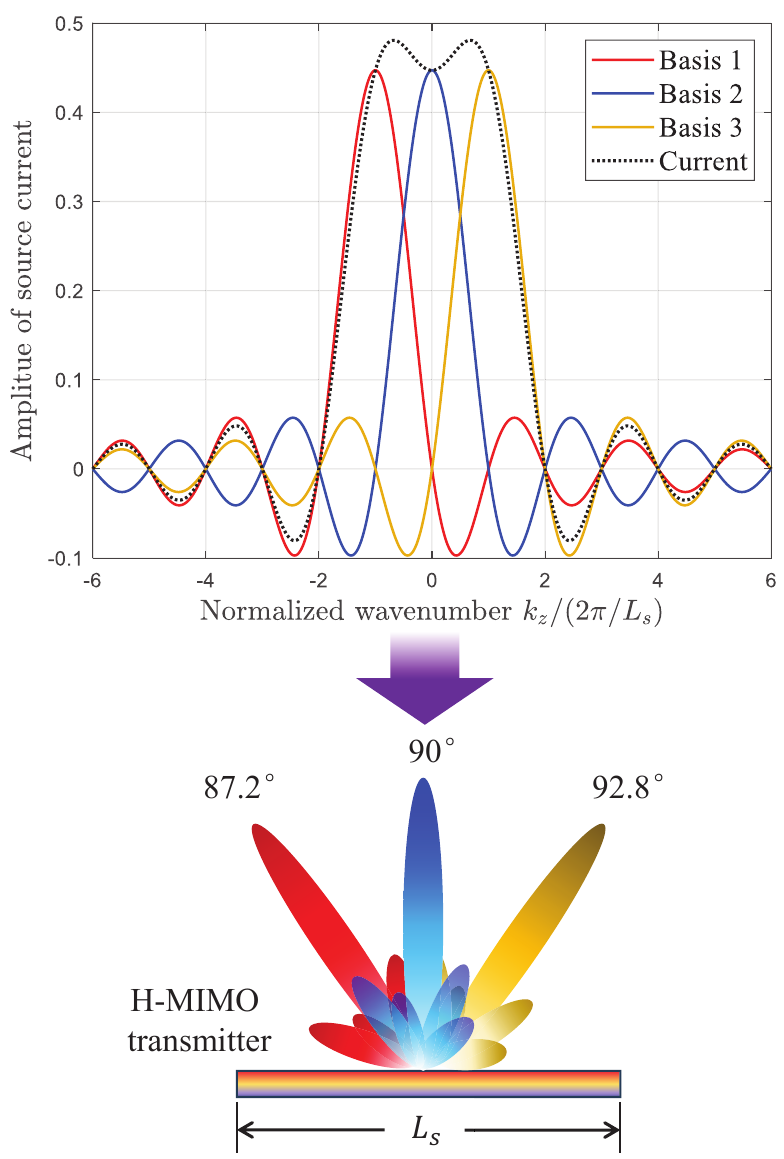}
	\caption{An illustration of WDM scheme \cite{Sanguinetti'23}. The source current is composed of three nearly orthogonal basis functions, each carrying one data stream.}
	\label{img:WDM}
\end{figure}

\begin{figure*}[!t]
	\centering
	\includegraphics[width=0.95\textwidth]{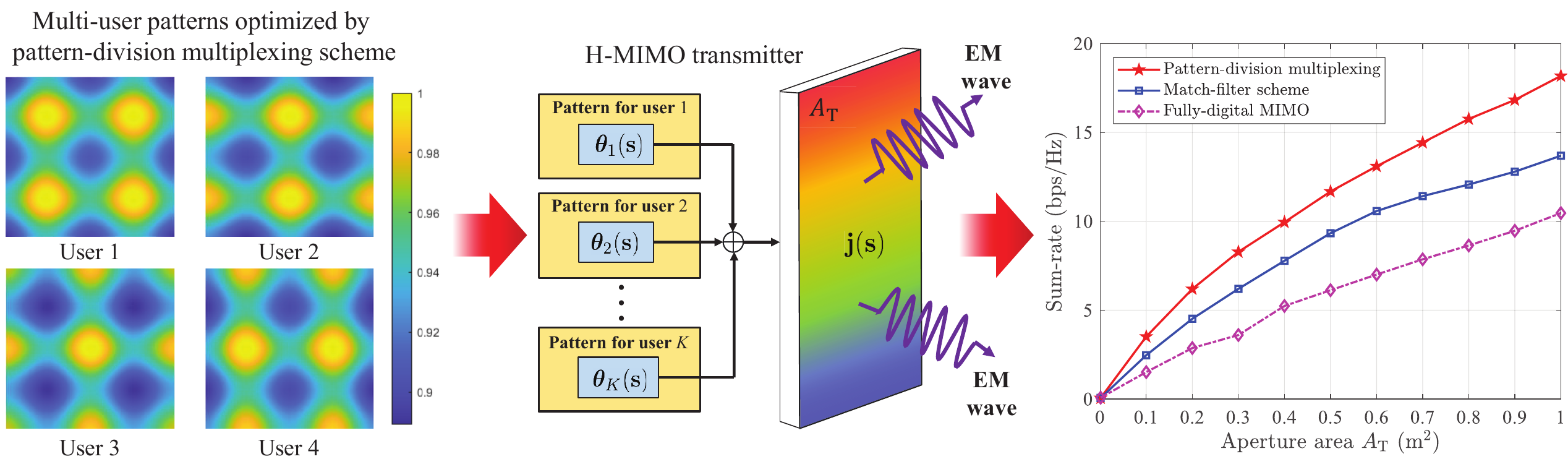}
	\caption{An illustration of multi-user H-MIMO downlink transmissions \cite{Zijian'23'JSAC}. The multi-user patterns are designed by PDM scheme for sum-rate maximization. Then, the optimized patterns are deployed on the holographic surface of H-MIMO transmitter to serve multiple receivers.}
	\label{img:PDM}
\end{figure*}

By representing continuous transmit currents and receive fields using a spatial Fourier basis, literature \cite{Sanguinetti'23} introduced a wavenumber-division multiplexing (WDM) scheme. The analyses have shown that transmitting multiple spatial sinusoids across the aperture yields parallel “communication modes,” whose number and quality depend on aperture size. In particular, when the receiver array size approaches the propagation range, the simplest digital WDM scheme can achieve the same sum-rate as the water-filling MIMO precoding. In practice, one can implement WDM by activating particular spatial patterns on a reconfigurable surface, which may be simpler than the optimal water-filling method.

Based on this idea, the WDM proposed in \cite{Sanguinetti'23} treats the continuous aperture much like an orthogonal frequency division multiplex (OFDM) system in space. As shown in Fig. \ref{img:WDM}, transmit waveforms are decomposed into orthogonal plane-wave modes (wavenumbers) via a spatial Fourier transform. Each mode carries an independent data stream and induces a waveform towards a different angle. The original WDM work focused on light-of-sight channels between transceivers, while one can also use a dictionary of Fourier-basis waveforms to multiplex users. Although simple to implement, WDM may suffer from inter-mode interference when the receiver aperture is not sufficiently large. In the future, the improvements of WDM include using optimized weights to mitigate the inter-mode interference.


\subsection{Wavenumber-Domain Channel Estimation}
As shown in Fig. \ref{img:channel}, H-MIMO channels exhibit clustered sparsity in the wavenumber domain, where only certain ranges of wavenumbers carry significant power. Techniques such as compressed sensing can exploit this sparsity. More generally, one can design wavenumber-domain probing waveforms (e.g. multi-sinusoidal spatial pilots) that target specific wavenumber bands, and then use sparse reconstruction to infer the channel. Different from classical compressed sensing, these designs use continuous wavenumber “probes” instead of a fixed DFT grid, thereby avoiding ``energy leakage''.

For example, the work \cite{guo2024wavenumber} formulates downlink training for an H-MIMO transmitter. It shows that the angular power spectrum of H-MIMO channels is continuous and cluster-sparse, making  DFT-based estimation inaccurate. Thus, the authors represent the channel as an integral over wavenumbers using a finite Fourier basis. This eliminates “energy leakage” from mismatched angles. The estimation problem becomes sparse recovery of the wavenumber-power distribution. Then, the authors proposed a graph-cut swap expansion (GCSE) algorithm that iteratively selects the most significant wavenumber modes. Simulations demonstrated that this estimator outperforms DFT-sampled angular-domain methods. 

\subsection{Wavenumber-Domain Waveform Design}
By exploiting near-field DoFs, H-MIMO can multiplex more users than traditional MIMO \cite{Cui'22'TCOM}. The wavenumber-domain waveform design plays a key role in this process. In a downlink system, a base station could assign disjoint sets of wavenumber modes to different data streams, effectively separating them in both angle and distance. Specifically, the set of Fourier basis functions can be treated as a wavenumber-domain waveform codebook. Particularly, compared to traditional DFT or polar-domain codebooks \cite{Cui'22'TCOM}, wavenumber-domain codebooks automatically adapt to near-field users. In a single-user H-MIMO link, one can select a Fourier-basis waveform that best matches the channel. In multi-user systems, a wavenumber-domain codebook can be combined with digital precoding to serve multiple users. To reduce the training overhead, codebooks can also be designed by quantizing the wavenumber domain, e.g., selecting a finite subset of Fourier modes that capture most channel energy.

For single-user systems, the authors in \cite{Yuanbin'25} introduced using wavenumber-domain bases for spatial waveform designs. In a single-user scenario, they compare a conventional DFT codebook against a wavenumber-domain codebook. The wavenumber codebook consists of steering vectors corresponding to plane waves across a finer angular grid. Remarkably, they find that the wavenumber codebook yields a rate that is nearly constant over distance, while the DFT codebook’s rate fluctuates and degrades as the user moves into the near field. This is because some DFT beams, especially the beams corresponding to high angles, have no physical counterpart when the user is close, causing those codewords to have negligible gain. By contrast, the wavenumber codewords are aligned with actual propagating modes. The wavenumber beams effectively eliminate power leakage, leading to better waveforming gains.

For multi-user systems, the wavenumber-domain waveform design has been studied in the literature. As shown in Fig. \ref{img:PDM}, the authors in \cite{Zijian'23'JSAC} modeled a multi-user H-MIMO system and developed a pattern-division multiplexing (PDM) scheme to design the patterns, i.e., the current-density distribution functions, for multiple users. The key idea is to transform the design of the spatial pattern functions to the design of their projection lengths onto finite Fourier bases in the wavenumber domain. Utilizing PDM, a block coordinate descent based pattern design scheme is proposed to maximize the sum-rate.  Fig. \ref{img:PDM} shows that, the optimized patterns are almost mutually orthogonal, which balances the directionality of multi-user beams and the elimination of inter-user interference. Simulations show the superiority of this wavenumber-domain design scheme over the existing spatial-domain schemes \cite{Zijian'23'JSAC}.

\subsection{Wavenumber-Domain Integrated Sensing and Communications}
H-MIMO has the potential to simultaneously support communication and perform environmental sensing tasks such as radar, imaging, and localization. In integrated sensing and communication (ISAC) systems, wavenumber-domain waveforms can be designed to form multiple simultaneous beams—some directed at communication users and others aimed at probing the environment. The unified wavenumber framework naturally accommodates this functionality, as various beam types can be expressed and processed in the wavenumber domain. Accordingly, certain spatial modes can be allocated for data modulation and transmission, while orthogonal modes are reserved for radar sensing. By correlating the received signals in the wavenumber domain, the sensing receiver can estimate distance–angle profiles with high fidelity.

The principles of wavenumber-domain designs can be effectively extended to waveform generation in ISAC systems. For instance, the authors in \cite{Xiangrong'25} investigated near-field ISAC in wideband communication scenarios. Due to the beam squint in both angle and distance, near-field beams are highly frequency-dependent, thereby rendering classical DFT codebooks ineffective for wideband waveform. This observation motivates the exploration of joint wavenumber–frequency ISAC design, wherein frequency-dependent wavenumber-domain waveformers can simultaneously direct energy toward users and illuminate sensing targets. Moreover, for radar applications, transmitting wideband wavenumber-domain pulses enables high-resolution distance–angle estimation. These techniques pave the way for truly “holographic” devices that use a shared electromagnetic aperture to perform both sensing and communication tasks efficiently \cite{HuangChongwen'20}.

\subsection{Hardware-Constrained Waveform Designs}
Practical H-MIMO arrays often face hardware limitations, such as the maximum power consumption and a limited number of RF chains. Design approaches in this category explicitly incorporate these hardware limitations into the wavenumber-domain waveformers. For instance, a desired wavenumber-domain waveform can be projected onto the feasible set defined by a reconfigurable intelligent surface (RIS) linked by RF chains \cite{Zijian'TST}. Employing the wavenumber basis within a hybrid precoding architecture is promising, as analog waveformer enabled by RISs can inherently implement a wavenumber transform with low power consumption and few RF chains. These hybrid analog-digital architectures can utilize a subset of wavenumber modes as analog beams, with low-dimensional digital precoding applied to refine the resulting signal. 

In certain scenarios, RIS-based H-MIMO systems may offer only phase-only control or limited amplitude tunability \cite{Zijian'TST}. However, accurate synthesis of wavenumber-domain waveforms typically requires precise control over both amplitude and phase across the aperture. With phase shifters alone, some wavenumber components may become inaccessible, resulting in suboptimal array patterns. A common practical approach involves solving a constrained optimization problem to identify the closest achievable pattern to realize the desired beam. Alternatively, amplitude-only designs in pattern-divided metasurfaces can be employed to implement selected Fourier modes with reduced hardware complexity \cite{Zijian'TST}.


\section{Open Challenges and Future Directions}

Despite their promises, the wavenumber-domain signal processing for H-MIMO is still in its infancy. In this section, we highlight several open challenges and future research directions.

\subsubsection{Capacity Theory}
While wavenumber-domain channel models provide a unified framework for both near- and far-field regimes, a comprehensive information-theoretic framework that fully incorporates EM properties remains undeveloped. Classical Shannon theory does not capture phenomena such as EM coupling at subwavelength scales or the behavior of evanescent modes. The emerging field of EIT seeks to bridge this gap by integrating Maxwell’s equations directly into capacity analyses \cite{Jieao'24}. Advancing such theory for H-MIMO systems will require a rigorous understanding of how wavenumber-domain sparsity and continuous aperture dimensions influence the DoFs and capacity limits. This constitutes a significant and open theoretical challenge in the field.


\subsubsection{Tri-Hybrid MIMO Architectures}
Even in classical massive MIMO deployments, introducing a layer of wavenumber-domain signal processing at the base station could enhance performance. For instance, to improve the energy efficiency, a massive array could apply an analog EM signal processing array (via lens or RIS) followed by a smaller hybrid precoding MIMO \cite{heath2025tri}. This tri-hybrid strategy allows to use a wavenumber-domain filter to reduce channel dimension while preserving capacity. It also reduces the number of required RF components and baseband processing units. Due to the increasing tension between larger bandwidths and infrastructure in 6G  systems, this architecture can offer a pathway to scale MIMO spatial dimensions without increasing the array aperture.


\subsubsection{Mutual Coupling Effects}
MC of H-MIMO poses both limitations and opportunities. From the challenge perspective, MC can distort the intended wavenumber-domain response and reduce array gains. In practice, this issue can be mitigated by 1) using metasurface-based apertures, where passive elements are naturally low-coupled; 2) employing decoupling structures such as neutralization lines, parasitic elements, metamaterial isolators; or 3) adopting coupling-aware calibration to reconstruct an effective wavenumber-domain basis. On the other hand, recent studies \cite{GuoYuqing'25} demonstrated that MC can enhance MIMO capacity in some scenarios. The modified radiation eigenmodes created by MC can enrich the spatial DoFs, enabling additional wavenumber-domain patterns. This effect may enhance focusing capability or provide additional diversity. Understanding how to exploit, rather than merely avoid, MC constitutes an important research direction for future H-MIMO signal processing.

\subsubsection{Robust Waveform Designs}
Real-world channels are imperfect, where scattering, mobility, and hardware impairments will deviate from ideal models. Designing waveforms that are robust to model mismatch is essential. A practical solution is to transmit a set of known spatial pilot patterns and measure the resulting fields across the aperture or at reference points in space. These measurements can be used to construct a series of effective bases, which can better reflect the true EM behavior of the aperture. Handling non-ideal scattering is another relevant challenge. One solution is to incorporate uncertainty sets around the dominant wavenumber support and design waveforms that are robust in the worst-case sense. These techniques complement wavenumber-domain modeling by accounting for environmental imperfections, thereby ensuring performance in practical systems.



\subsubsection{Network-Level Integration}
From a network-level perspective, wavenumber-domain processing should also align with system-level constraints. Performing wavenumber-domain filtering at the base station introduces additional computational steps beyond conventional DFT-sampled angular-domain processing. To remain scalable, practical systems may employ reduced-order wavenumber-domain models or hierarchical transforms that limit complexity while preserving the dominant spatial modes. Moreover, 6G pre-standardization discussions increasingly consider near-field measurement reference signals, multi-antenna feedback compression, and distributed MIMO architectures. To enable  compatibility, wavenumber-domain representations should support these developments by providing a unified signal processing framework.




\section{Conclusion}
In this article, we have overviewed the foundations and state-of-the-art of wavenumber-domain signal processing for H-MIMO. We started by introducing H-MIMO’s key characteristics and the limitations of traditional models, motivating the need for a Fourier plane-wave perspective. We covered the core background on H-MIMO channel models and wavenumber representations. We then classified different design methodologies and summarized recent research contributions in multiplexing schemes, channel estimation, waveform designs, and so on. Finally, we highlighted open challenges, such as capacity theory and wideband communications, which present some opportunities for future works.

In the future, wavenumber-domain signal processing is expected to play an increasingly important role in 6G standardization. As global standardization bodies begin considering near-field channel models and extremely large-scale apertures, a unified wavenumber-domain formulation offers a physically-consistent basis for defining channel representations and codebooks. Moreover, wavenumber-domain sparsity and spatial bandwidth concepts provide new tools for specifying channel feedback mechanisms, aperture-domain reference signals, and EM-aware performance metrics suitable for H-MIMO. In practice, constructing and processing a continuous wavenumber spectrum introduce additional computational overhead and require more sophisticated calibration to handle model mismatch or hardware imperfections. The implementation complexities, including sampling density, power consumption, and real-time processing capability, should be carefully balanced against the performance gains.

\section*{Acknowledgment}
This work was supported in part by the National Natural Science Foundation of China (Grant No. 62325106), in part by the National Natural Science Foundation of China (Grant No. 62031019), and in part by the National Key Research and Development Program of China (Grant No. 2023YFB3811503).



\bibliographystyle{IEEEtran}
\bibliography{IEEEabrv,reference}


\section*{Biographies}
\textbf{Zijian Zhang} received his Ph. D. degree in Tsinghua University  in 2025. His research area is beyond massive MIMO for 6G wireless communications. He has received the Tsinghua Sci. Technol. Best Paper Award in 2024 and the IEEE SPS Young Author Best Paper Award in 2025.
\\

\textbf{Linglong Dai} is a Professor from Tsinghua University. His current research interests include massive MIMO, RIS, Wireless AI, near-field communications, EIT, and quantum information. He has received the IEEE ComSoc Leonard G. Abraham Prize in 2020, the IEEE ICC 2022 Outstanding Demo Award, the National Science Foundation for Distinguished Young Scholars in 2023, and the IEEE ComSoc Stephen O. Rice Prize in 2025. He was listed as a Highly Cited Researcher by Clarivate from 2020 to 2025. He was elevated as an IEEE Fellow in 2021.
\end{document}